\documentclass[twocolumn,epsfig,pre]{revtex4}

\usepackage{graphics}
\usepackage{graphicx}
\usepackage{epstopdf}
\usepackage{amssymb}
\usepackage{amsmath}
\usepackage{hyperref}
\usepackage{latexsym}
\usepackage{color}

\begin{document}
	
	\renewcommand{\thefootnote}{\fnsymbol{footnote}}
	\renewcommand{\theequation}{\arabic{section}.\arabic{equation}}
	
	\title{Effect of topology on the collapse transition and the instantaneous shape of a model heteropolymer}

	\author{Thoudam Vilip Singh}
	\email{thoudamsinghsingh@gmail.com}	
	\author{Lenin S. Shagolsem}
	\email{slenin2001@gmail.com}
	\affiliation{Department of Physics, National Institute of Technology Manipur, Imphal, India} 
	
	\date{\today}

\begin{abstract}

\noindent The effect of topology on the collapse transition and instantaneous shape of an energy polydisperse polymer (a model heteropolymer) is studied by means of computer simulations. In particular, we consider three different chain topology, namely, linear (L), ring (R) and trefoil knot (T). The heteropolymer is modeled by assigning each monomer an interaction parameter, $\varepsilon_i$, drawn randomly from a Gaussian distribution. Through chain size scaling the transition temperature, $\theta$, is located and compared among the chains of different topogies. The influence of topology is reflected in the value of $\theta$ and observed that $\theta(\text{L}) > \theta(\text{R}) > \theta(\text{T})$ in a similar fashion to that of the homopolymer counterpart. Also studied chain size distributions, and the shape changes across the transition temperature characterised through shape parameters based on the eigenvalues of the gyration tensor. It is observed that, for the model heteropolymer, in addition to chain topology the $\theta$-temperature also depends on energy polydispersity. 

\end{abstract}

\maketitle

%%\tableofcontents

%%%%%%%%%%%%%%%%%%%%%%%%%%%%%%%%%%%%%%%%%%%%%%%%%%%%%%%%%%%%%%%%%%%%%
\section{Introduction}
\label{sec: intro}

Polymers exist in a multitude of architecture and shapes. The topology of polymer chains plays a crucial role in modifying physical properties such as strength, toughness, glass transition, and mechanical response, leading to interesting applications. For example, polypropylene, a linear thermoplastic chain, is among the most successful comodity synthetic polymers, while polyester-based star polymers have excellent biocompatibility, biodegradability and posses properties suitable for advanced polymer therapeutics.\cite{Karger-Kocsis,Ren} Network polymers find applications in solar cells, antimicrobial coating and other range of applications.\cite{DeKeer} On the other hand, mixing of different topology, e.g.~linear and ring polymers, is one way to alter physical characteristics of the polymeric material. In this regard, ring polymer, a topologically constrained chain, is relevant not only in physics, but also in biology as a model system for chromatin folding and existence of chromosome territories. \cite{mcleish2002,polymerpoulous207,halverson2014,shagolsem2018} \\
\medskip

The existance of circular topology in biopolymer, e.g. in DNA of bacteriophage, $\phi{\rm X174}$, was observed as early as 1964.\cite{Freifelder} More complicated topologies such as knotted and concatenated DNAs also play an important role in recombination and replication (with the help of topoisomerases).\cite{Wasserman1,Champoux} A knot is a three dimensional topological state of a closed and non-intersecting curve, while its absence is referred to as unknot or trivial knot. These knots are not restricted to bacteria and virus, but are found in other biophysical systems.\cite{Meluzzi} Although the presence of knots drastically weakens the biomolecule, this may result in the increase in stability of some systems, which could explain the presence of knots in certain proteins.\cite{Yeates} In biological systems, knots are formed via threading of loose ends (e.g., random cyclization in linear double-stranded DNA), \cite{Rybekov,Shaw} and breaking and rejoining of segments assisted by enzymes like recombinases. \cite{Buck,Lui,Stark,Wasserman}
%Trefoil knots in nanoscales have been tied artificially using optical tweezers. 
%\medskip
Mathematically, a trefoil is the simplest torus knot, i.e., it can be drawn on the surface of a doughnut-shaped object. Knot theory have been used to elucidate site-specific recombination of DNAs.\cite{Ernst,Shimokawa} Also, many unique properties of DNA exhibiting topologies like cyclic, knot or even links (catenases) are explained satisfactorily using topological geometry and graph theory. Such chains are vital for understanding the evolution of variety of biological functions.\cite{Shimokawa,Seeman} In spite of great progress in the study of chains of various topologies, there are many things left to be understood. For instance, the transition of knots from one type to the other in both open and closed chains are far from understood. Theoretical and computational studies have tremendous promise in understanding the knotting mechanism which could lead to useful applications of molecular knots in bionanotechnology, bionanomedicine and biotechnology. 
%The probability of finding a knot of any type $X$ in an $n$-step self avoiding random walk is $P\left(X\right) \sim \text{exp}\left(–n/n_0\right)$, where $n_0$ is a model dependent constant and the prefactor being dependent on the knot type. For a polymer chain obeying self avoiding random walk with $N$ monomers, its radius of gyration $R_g \sim N^{\nu}$ where $\nu$ ≈ 0.588 is scaling exponent \cite{Janse}. With the help of Monte Carlo simulation, the same has been observed for circular polymers that are both knotted and unknotted provided N tends to $\infty$ \cite{Grosberg}. Measurements on 31, 41, 51, 52, and 71 knots in dsDNA showed knot lengths of 250-550 nm for molecules stretched due to tension of about 1 pN \cite{Bao}. Tight knots are also found in proteins and knowing their size is of great importance in comprehending the biological roles of proteins.

Since the polymer chains can exists in different topologies its shape, size and conformational properties are also varied. Studies on ring polymer melts show that topological constraints influence both statics and dynamics. For example, sufficiently large ring polymers behave as a globular object, where its size $\sim N^{1/3}$ with $N$ chain length, unexpected power-law stress relaxation in entangled ring polymers, higher diffusion coefficient or lower viscosity in comparison with the linear counterpart.\cite{kapnistos2008,richter2015} These studies clearly illustrates the relationship between the chain topology and its consequences on the physical properties of the material. On the other hand, the shape of a protein can determine its function, while for DNA it is important in protein-DNA recognition.\cite{Rohs} Hence, the study of polymer morphologies or shape parameters are quite important. 
It is known that for linear polymers, the chain size $R_g \sim N^{\nu}$, where the Flory exponent $\nu = 1/2$ (ideal chain or chain in $\theta$-solvent).\cite{Steinhauser} While for polymer with knots $R_g \sim N^{\nu}C^{1/3 - \nu}$ with $C$ the number of essential crossing.\cite{Quake} Here, for self-avoiding trivial knots (unknotted rings) $\nu=3/5$ for large $N$. 
A better expression for knots that describes the dependence of $R_g$ on $N$, temperature and topology uses a weak topological invariant of knots known as $P$-parameter, which is the aspect ratio of chain length to the diameter of the knotted polymer in a maximally inflated tube and the results at various regimes are summarised in table~\ref{table: Rg}.\cite{Grosbergg} \\
%\begin{equation}
%R_g =
%\begin{cases}
%bN^{3/5}\zeta^{1/5}P^{-4/15}, \text{in good solvent regime}\\
%bN^{1/2}P^{-1/6}, \text{in quasi-Gaussian regime} \\
%bN^{1/3}|\zeta|^{-1/3}\left[1+|\zeta|^{-4/3}\left(\frac{P}{N}\right)^{2/3}\right],\\ 
%\qquad \qquad \qquad  \quad \quad \text{in poor solvent regime}\\
%bN^{1/3}, \text{in maximally tightned knot regime} 
%\end{cases}
%\end{equation} 

\begin{table}
	\begin{center}
		\caption{Dependence of $R_g$ on temperature, $N$, and topology for knots. Here, $b$ is the monomer size, and $\zeta = 1 - T/T_{\theta}$ is a dimensionless deviation from $\theta$.}
		\begin{tabular}{ c  c }
			\hline
			$R_g $	&	Regime	\\ 		\hline
			$ bN^{3/5}\zeta^{1/5}P^{-4/15}$	&	good solvent 	\\ 
			$bN^{1/2}P^{-1/6}$				&	quasi-Gaussian 	\\ 
			$bN^{1/3}|\zeta|^{-1/3}\left[1+|\zeta|^{-4/3}\left(\frac{P}{N}\right)^{2/3}\right]$	&	poor solvent 	\\
			$bN^{1/3}$						&	maximally tightned knot	\\ \hline
		\end{tabular}
		\label{table: Rg}
	\end{center}
\end{table}

\medskip

In the present work, the effect of topology on the instantaneous shape of an energy polydisperse polymer (EPP), a model heteropolymer system in the limit where the number of species is equal to the total number of monomers, is investigated by means of molecular dynamics simulations. 
The EPP chain also represents a diordered system which is often used as a statistical model to study proteins/DNA (or intrinsically disordered proteins in general) at coarse-grained level.\cite{Pande,Shakhnovich} Moreover, biopolymers exists in complex topological states tailored for specific functions and thus understanding the interplay between topology and collapse transition and instantaneous shape of the model heteropolymer is of interest. 
In EPP chain, each monomer has unique identity characterized by its interaction parameter drawn randomly from a given distribution (detailed in next section). In a recent work,\cite{vilip2022} two types of linear EPP chains with different energy distribution functions (i.e., Gaussian and uniform) were considered, where the focus was on the effect of functional form and variance of the energy distribution on the nature of collapse transition in general. It was observed that, the transition temperature depends of the type of distribution and it is governed by the most probable value of the distribution rather than the width of the distribution. On the other hand, the nature of collapse transition is universal, i.e. independent of the functional form and variance under proper scaling. However, in the current study, we focus on the role of chain topology by considering EPP chain with Gaussian distribution function.
In particular, we examine the details of shape changes during coil-globule transition of two relatively simple topology, namely, ring (R) and trefoil knot (T) which are present in biological systems also, and linear (L) chain as the reference system.  
\medskip

Our paper is organized as follows. Model and simulation details are described in section~\ref{sec: model-description} followed by the results in section~\ref{sec: results}, where the effect of topolgy on $\theta$-temperature, instantaneous shapes, chain size dependence and distributions are discussed in sub-sections~\ref{sec: theta-temp}--\ref{sec: probability-distribution}, and finally conclude in section~\ref{sec: summary}. 

%\clearpage

%%%%%%%%%%%%%%%%%%%%%%%%%%%%%%%%%%%%%%%%%%%%%%%%%%%%%%%
%%%%%%%%%%%%%%%%%%%%%%%%%%%%%%%%%%%%%%%%%%%%%%%%%%%%%%%

\section{Model and simulation details}
\label{sec: model-description}

We employ coarse grain bead-spring model of Kremer-Grest,\cite{Grest} to simulate the EPP chain. Here, the beads representing coarse-grained monomers interacts via pair-wise Lennard-Jones (LJ) potential which, between $i-j$ monomers pair, is given by 
\begin{equation}
U_{\tiny{_{\rm LJ}}}(r) = 4\varepsilon_{ij}\left[(\sigma/r)^{12}-(\sigma/r)^{6}\right],
\label{eqn: LJ-potential}
\end{equation}
where the pair-wise interaction strength $\varepsilon_{ij} = \sqrt{\varepsilon_i \varepsilon_j}$ (following Lorentz-Berthelot mixing rule \cite{Berthelot}), $\sigma$ the effective monomer diameter, and $r$ the separation between the monomers. The potential is cut-off and shifted to zero at $r_c=2.5\sigma$. The monomer connectivity along the chain is modeled using the finitely extensible, nonlinear, elastic (FENE) potential defined as 
\begin{equation}
U_{\tiny{_\text{FENE}}} = \left\{
\begin{array}{l l}
-\frac{kr_0^2}{2}\ln\left[1-\left(r/r_0\right)^{2}\right]~, & \quad r<r_0 \\ %\nonumber\\ 
\infty~, & \quad r\ge r_0
\end{array} \right.
\label{eqn: fene-potential}
\end{equation} 
where $r_0$ = 1.5$\sigma$ is the  maximum extension along the chain, and the value of spring constant is taken to be $k = 30\varepsilon/\sigma^2$.\cite{Grest,Kremer} %These values ensures that bond crossing and very high frequency modes are avoided. 
%\medskip
The EPP chain is constructed in the following way. To every monomer we assigned an interaction energy, $\varepsilon_i$, drawn randomly from a Gaussian distribution 
\begin{equation}
P(\varepsilon_i)= 
\frac{1}{\sqrt{2 \pi \left( \text{SD} \right)^2}} \text{exp} \left[-\frac{\left(\varepsilon_i - \langle\varepsilon \rangle \right)^2}{4 \left( \text{SD} \right)^2} \right],
\label{eq: Gaussian distribution}
\end{equation} 
where $\langle \varepsilon \rangle$ and SD are the mean and standard deviation respectively. The mean of the energy distribution is fixed at $\langle \varepsilon \rangle = 2.5$ for all systems, and the value of SD varies in the range $0.01-0.10$ corresponding to variation of polydispersity index, $\delta$ = SD/mean, in the range $4\%-13\%$. (Use of other functional form, e.g.~log-normal distribution at small value of polydispersity index, to model heteropolymer is reported in reference~\cite{Vilip2021}.) A homopolymer (HP) chain with the interaction energy equal to the mean of the distribution is taken as our reference system. 
All the physical quantities in this study are expressed in LJ reduced units,\cite{Allen, Frenkel} in which $\varepsilon$ and $\sigma$ are the energy and length scales respectively. The reduced temperature, $T= k_{\rm{B}}T_0/\varepsilon$, and time, $t=t_0/\tau_{\text{LJ}}$ with $\tau_{\text{LJ}} = \sigma(m/\varepsilon)^{1/2}$ the LJ time, and $k_{\text{B}}$, $T_0$, $t_0$ and $m$ are the Boltzmann constant, absolute temperature, real time and real mass respectively. The diameters and the mass of all the monomers are fixed at $\sigma = 1$ and $m = 1$ respectively. 
\medskip

\begin{figure*}[ht!]
	\begin{center}
		\includegraphics[width=0.97\textwidth]{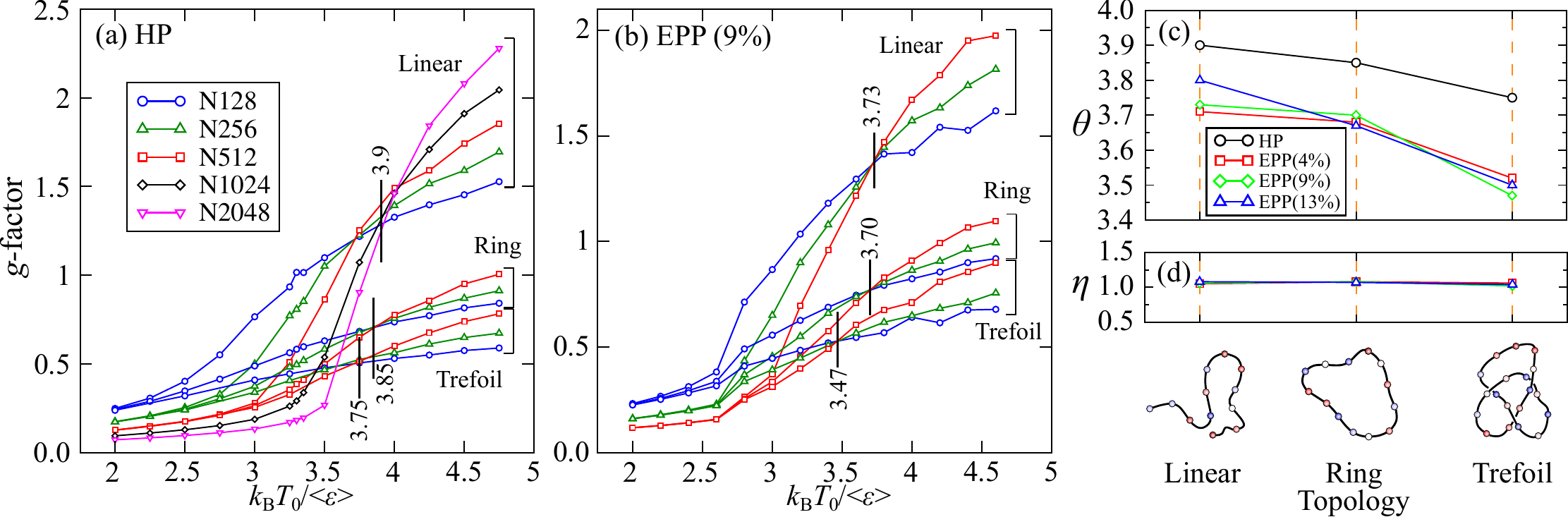} 
		\caption{The plot of $g$-factor vs $T$ for (a) homopolymers and (b) EPP chains of different topology indicated in the figure at $\delta$ = 9\%. (c) The values of $\theta$-temperature for linear, ring, and trefoil knot are shown for homopolymer (HP) and EPP at different values of $\delta$. (d) $\eta$ represents the ratio of \textit{g}-factor at $\theta$ for EPP to that of the homopolymer of same topology. Its values remain approximately unity for liear, ring and trefoil knots.}
		\label{fig: xi-vs-T}	
	\end{center} 
\end{figure*}

We use molecular dynamics (MD) simulations using Langevin dynamics \cite{Allen, Frenkel} where the equation of motion is given by
\begin{equation}
m_i \frac{\text{d}^2 {\bf r}_i}{\text{d}t^2} + \gamma \frac{\text{d} {\bf r}_i}{\text{d}t} = -\frac{\partial U}{\partial {\bf r}_i} + {\bf f}_i(t)~,
\label{eqn: langevin}
\end{equation} 
in which ${\bf r}_i$ and $m_i$ are the position and the mass of particle $i$, respectively. $\gamma$ is the friction coefficient which is  the same for all particles, and $U = U_{\tiny{_{\rm LJ}}}+U_{\tiny{_\text{FENE}}}$ is the potential acting on monomer $i$ of the polymer chain, and ${\bf f}_i$ represents random external force which follows the relations: $\left\langle {{\bf f}_i}(t)\right\rangle=0$ and $\left\langle {{\bf f}_i}^\alpha(t){{\bf f}_j}^\beta(t')\right\rangle=2\gamma m_i k_{_{\rm B}}T\delta_{ij}\delta_{\alpha\beta}\delta(t-t')$ where $\alpha$ and $\beta$ denote the Cartesian components. 
The friction coefficient is $\gamma=1/\tau_d$, with $\tau_d$ as the characteristic viscous damping time which we fix at $50$ and it determines the transition from inertial to overdamped motion. The chosen value of $\gamma$ gives the correct thermalization for our investigation. The equations of motion are integrated using velocity-Verlet scheme with a time step of $\delta t_0 = 0.005 \tau_{_{\rm LJ}}$. \\
\medskip

A polymer chain of given topology is tethered at the centre of simulation box (of dimensions $L_x = L_y = L_z = 100\sigma$) which is periodic in all directions, and we consider chain length $N$ in the range $128-512$. Typically, at a given temperature, the chains are initially relaxed for $1 \times 10^6$ MD steps followed by production run of $5 \times 10^6$ MD steps where various measurements are performed. Since $\varepsilon_i$ are randomly distributed along the chain we consider 10 replicas (for a given value of mean and variance) and, for a faithful representation, the results reported here are the average taken from these replicas. All the simulations are carried out using LAMMPS code. \cite{Plimpton} 

%%%%%%%%%%%%%%%%%%%%%%%%%%%%%%%%%%%%%%%%%%%%%%%%%%
%%%%%%%%%%%%%%%%%%%%%%%%%%%%%%%%%%%%%%%%%%%%%%%%%%

\section{Results}
\label{sec: results}

\subsection{Effect of topology on $\theta$-temperature}
\label{sec: theta-temp}

Polymer chain, irrespective of topoloy, undergoes conformational transition, i.e. coil to globule, upon lowering temperature (or in going from good solvent to poor solvent condition). It is clear by now that the balance of net monomer-monomer interaction and the conformational entropy drives such collapse transition.\cite{Grosberg_book,Cates} 
%DISCUSSION ON TOPOLOGICAL EFFECT ON CHAIN SIZE, etc.. see Cates Duetch \cite{Cates}, etc...
%\medskip

Numerically, the transition temperature is located through $g$-factor,\cite{Zimm} which is defined as the ratio of the squared radius of gyration, $R_g$, of a topological chain to that of an ideal linear chain, i.e., 
\begin{equation}
g{\rm-factor}(N, T) = \frac{\langle R_g^2 \rangle (N, T)}{\langle R_g^2 \rangle_{_{\rm L}}(N, \theta)},
\end{equation} 
with $N$ as the number of monomers in a chain, $\theta$ denotes the $\theta$-temperature (i.e. temperature at which collapse transition occurs), $\langle R_g^2 \rangle_{_{\rm L}}(N, \theta)$ the squared $R_g$ for a linear chain of $N$ monomers at $\theta$, and angular brackets represent the ensemble average. 

As we can see in figure~\ref{fig: xi-vs-T}(a) and (b), all the curves of a particular topology meet at $\theta$-temperature for homopolymers and EPP chains respectively, where the ideal chain statistics is followed. For homopolymers, $\theta\approx$ 3.9 (L), 3.85 (R), 3.75(T), and for energy polydisperse polymers, $\theta\approx$ 3.73 (L), 3.70 (R), 3.47(T). In consistent with the earlier studies on topological effects,\cite{Suzuki, Narros} here also it is observed that the chain size decreases with increasing complexity of the topology, i.e $R_g({\rm L}) > R_g({\rm R}) > R_g({\rm T})$, also $\theta$-temperature follows the same trend $\theta({\rm L}) > \theta({\rm R}) > \theta({\rm T})$. This effect of topology on $\theta$-temperatures at different polydispersity index are summarised in figure~\ref{fig: xi-vs-T}(c). 

%\medskip

These are not only based on the minimization of free energy due to the contributions of energetic interations in the mean field approximation, and the entropic loss due to swelling of the chain, but also on the topological constraints. In unknotted and knotted rings, there exists a short range repulsion between different sections of the chain, and obviously it becomes more prominant in the knotted one. [Ref: Cates \& Duetch \cite{Cates}, Grossberg \& Rabin]

As the temperature is lowered below $\theta$, the decrease in global conformational entropy seems more in knots (followed by rings) than in linear chains, and it dominates over the local short range repulsion easily. Hence, $\theta$ is the lowest for trefoil knots. 
The homopolymer counterparts have larger values of $\langle R_g^2\rangle$ and $g$-factor, and thus correspondingly greater values of $\theta$. At a certain temperature, a trefoil would experience a better solvent condition than a ring with same $N$, which is again better than the corresponding linear chain.

\subsection{Instantaneous Shape Analysis} 
\label{sec: shape-analysis} 

\begin{figure}[ht!]
	\begin{center}
		\includegraphics[width=0.47\textwidth]{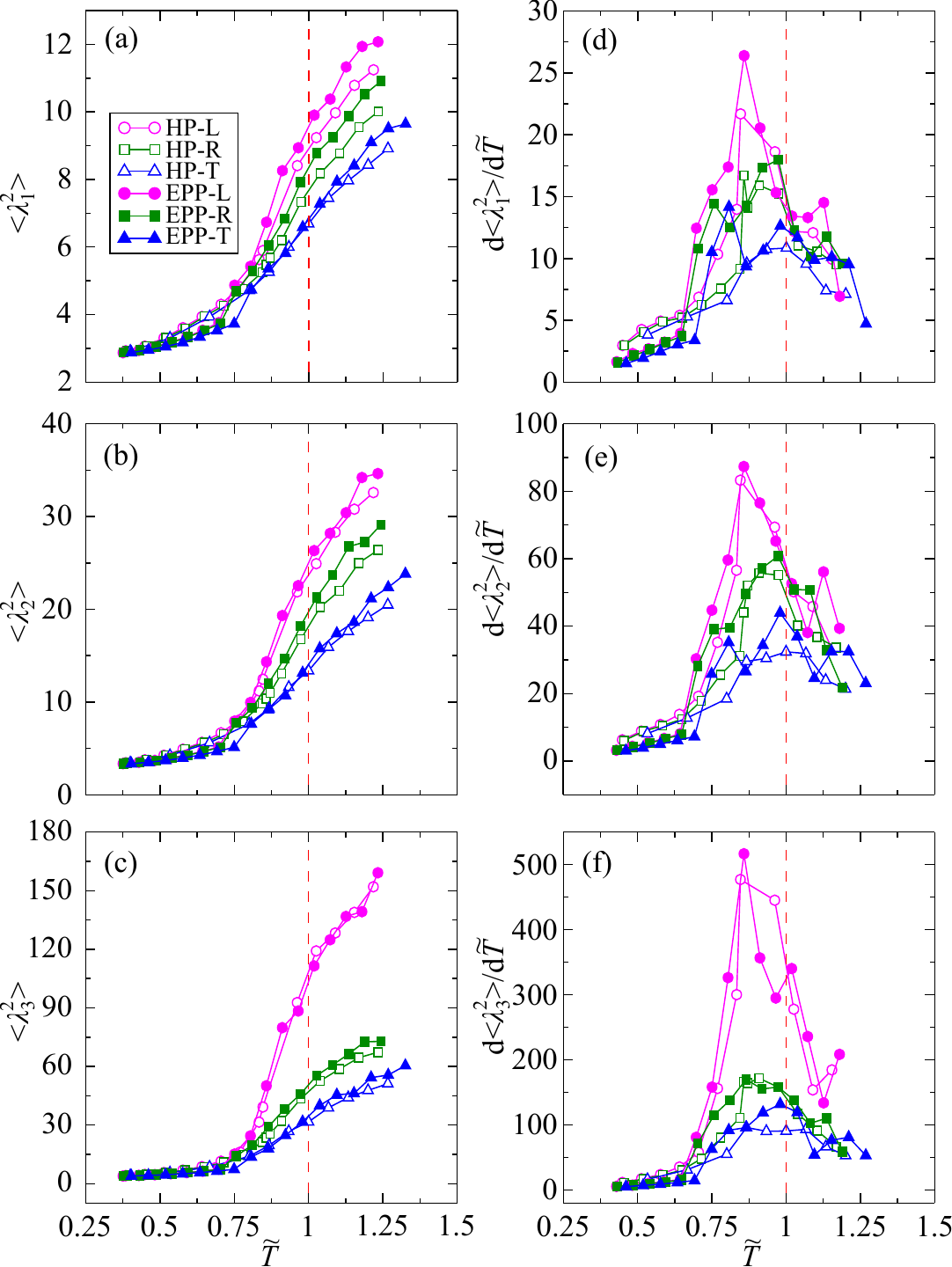}
		\caption{The dependence of average eigenvalues $\langle \lambda_1^2 \rangle$, $\langle \lambda_2^2 \rangle$, $ \langle \lambda_3^2 \rangle$ of gyration tensor on reduced temperature $\tilde{T}$ at $\delta=9\%$ are shown in figure (a)-(c). Their respective temperature derivatives are shown in (d)-(f).}
		\label{fig: lambda}	
	\end{center}
\end{figure}

The effect of topology on the shape changes of EPP chains during coil-globule transition is investigated through the eigenvalues of gyration tensor, {\bf S}, as a shape measure.\cite{Solc, solc_stockmayer_1971, solc1973} Here, by choosing a principal axis system in which {\bf S} is in diagonal form such that 
\begin{equation}
\langle R_g^2 \rangle = {\rm Tr({\bf S})} = \text{diag}\left(\lambda_1^2,\lambda_2^2,\lambda_3^2 \right), 
\end{equation} 
where $\lambda_1^2 \le \lambda_2^2 \le \lambda_3^2$ are the eigenvalues of {\bf S}. 
%These values represent the principle moments of the equivalent ellipsoid, and various shape parameters can be derived using the gyration tensor.
To this end we obtain average eigenvalues of the gyration tensor from 10 replicas (with 1000 independent configurations from each replica) for EPP chains. The eigenvalues obtained at different values of reduced temperature, $T^\ast=T/\theta$, are shown in figure~\ref{fig: lambda}(a)-(c) along with the corresponding change with respect to $T^\ast$, i.e. ${\rm d}\lambda_i/{\rm d}T^\ast$, in figure~\ref{fig: lambda}(d)-(f) for both EPP and HP chains of different topologies. 

%In figure~\ref{fig: lambda}, the eigenvalues of gyration and their fluctuations are shown as a function of reduced temperature, $\tilde{T} = T/\theta$ for EPPs with different topologies at $\delta$ = 9\% along with their homopolmer counterparts. 
At high temperature, all the eigenvalues $ \langle \lambda_1^2 \rangle$, $ \langle \lambda_2^2 \rangle$, $ \langle \lambda_3^2 \rangle$ of linear chains are relatively the largest, while that of the trefoil chains are the smallest. Furthermore, homopolymer counterparts of the same topology have lower values. These is still true at the transition point, $\tilde{T}$ = 1. As the temperature is lowered, all the curves seem to merge at $\tilde{T} \approx 0.65$. A clearer picture is seen in the derivative of $\tilde{T}$. At $\tilde{T} \approx 0.65$, the EEP as well as the homopolymer curves merge. The same for $\delta$ = 4 and 13\% can be seen in SI.

\begin{figure}[ht!]
	\begin{center}
		\includegraphics[width=0.47\textwidth]{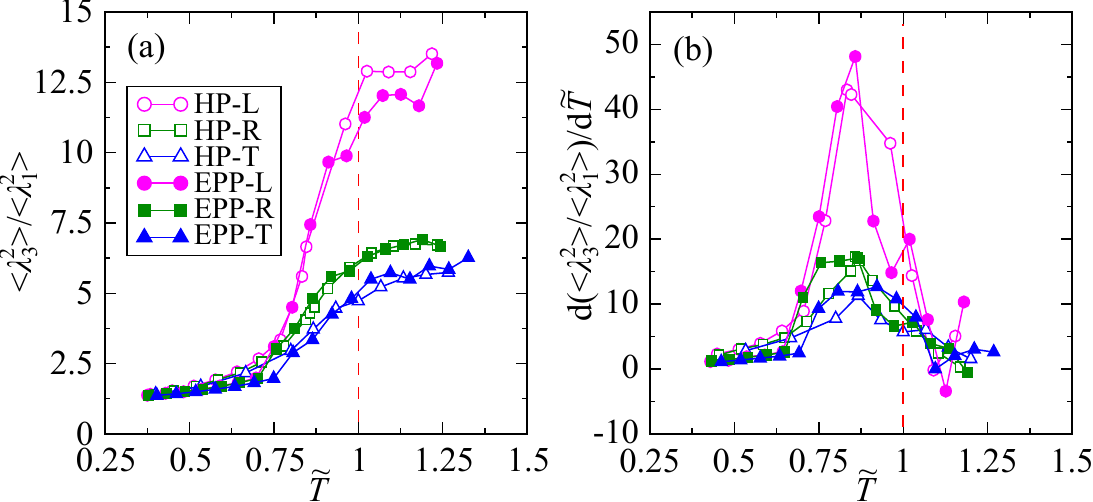}
		\caption{The plot of (a) $\langle \lambda_3^2 \rangle/ \langle \lambda_1^2 \rangle$ as a function of $\tilde{T}$ and (b) its temperature derivative for EPP chains at $\delta$ = 9\%.} 
		\label{fig: l3byl1}	
	\end{center}
\end{figure}

To support the above findings, the ratio of the largest eigenvalue to that of the smallest, $ \langle \lambda_3^2 \rangle/ \langle \lambda_1^2 \rangle$, and its temperature derivative are also plotted as a function of $\tilde{T}$ at $\delta$ = 9\%, as displayed in figure~\ref{fig: l3byl1}. The value of this ratio jumps to higher values at temperature just above 0.65, which is even more pronounced in the temperature derivative plot. At high temperature, the linear chains possess higher values of $ \langle \lambda_3^2 \rangle/ \langle \lambda_1^2 \rangle$ followed by the rings and finally the trefoils. 

\begin{figure}[ht!]
	\begin{center}
		\includegraphics[width=0.47\textwidth]{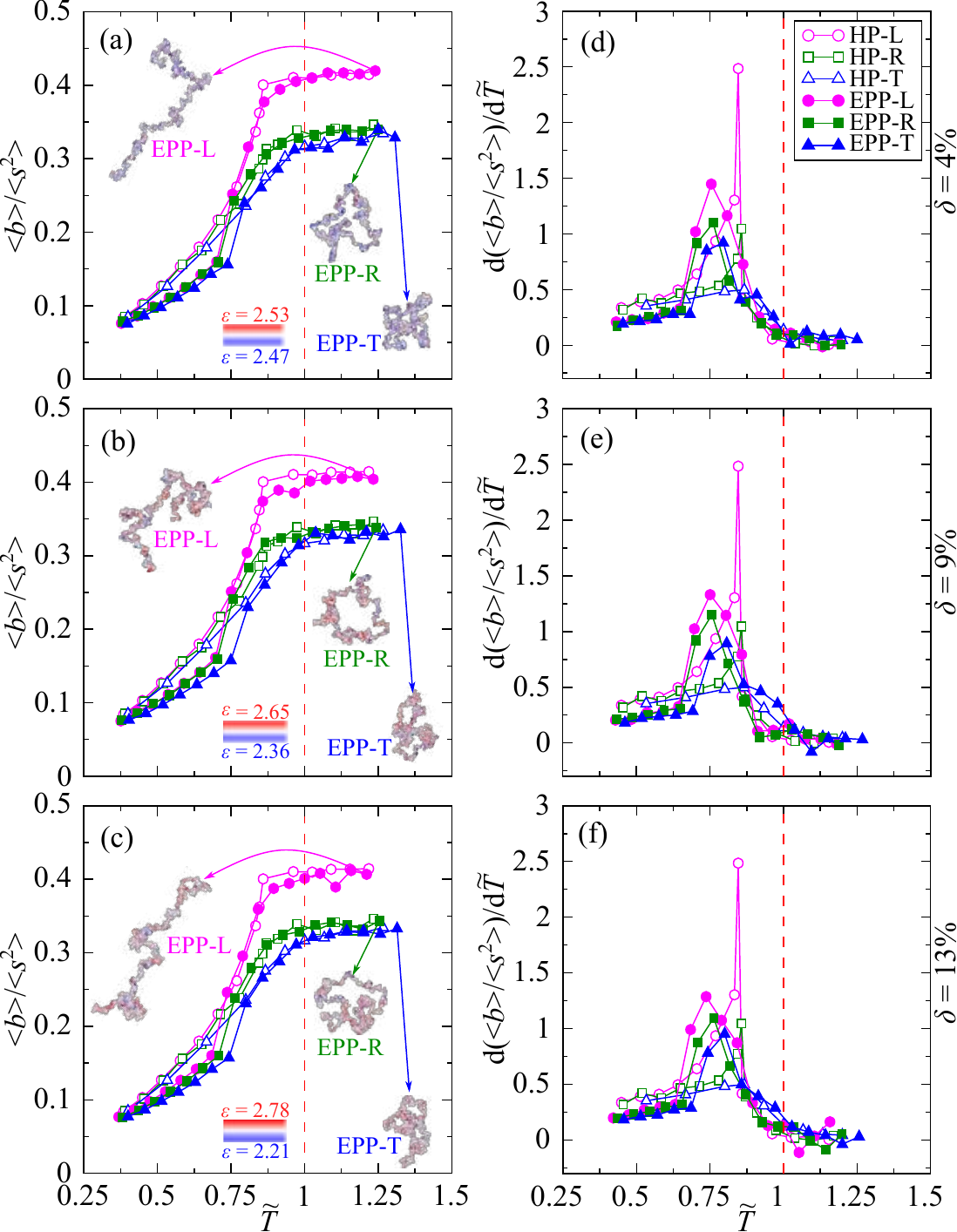}
		\caption{The variation of asphericity, $\langle b \rangle/\langle s^2 \rangle$ with $\tilde{T}$ at $\delta$ = 4, 9, and 13\% for different topology in figure (a)-(c). Their respective temperature derivaties are presented in (d)-(f).}
		\label{fig: b}	
	\end{center}
\end{figure}

The instantaneous shapes of the polymer chains are characterised through asphericity, defined as 
\begin{equation}
\langle b \rangle = \langle\lambda_3^2 \rangle - \frac{1}{2}\left(\langle\lambda_1^2 \rangle + \langle\lambda_2^2 \rangle \right).
\label{eq: b}
\end{equation} It checks how spherical the conformation is, and $b$ = 0 for a complete sphere. Its maximum is 1 for a completely collinear conformation. Since $b$ can be a linear combination of the average eigenvalues as in equation~\ref{eq: b}, curves similar to that in figure~\ref{fig: lambda} and \ref{fig: l3byl1} are observed. In the following figure~\ref{fig: b}, the normalised asphericity and their reduced temperature derivative as a function of $\tilde{T}$ for different topology at $\delta$ = 4, 9 and 13\% are presented along with that of the  homopolymers as reference. At high temperatures, all the curves saturate at $\langle b \rangle/\langle s^2 \rangle \approx$ 0.41 and 0.40 for linear homopolymer and EPP chains respectively, while it is 0.33 for both types of rings and trefoil knots. Their values at $\tilde{T} = 1$ for various topologies are summarised at table~\ref{table: b1}. When the temperatures are lowered, all polymer conformations tend to have a greater spherical symmetry. 

Trefoils have comparatively lower values of $\langle b \rangle/\langle s^2 \rangle $, indicating more spherical symmetry, which is followed by the rings. As temperature decreases below $\tilde{T} \approx 0.75$, homopolymer curves and the EPPs curves merge together. In general, with increase in polydispersity, the chains tend to be more spherical although this effect is quite small for rings and trefoils.

\begin{table}
	\begin{center}
		\caption{Different values of shape parameters at $\tilde{T} = 1$}
		\begin{tabular}{ c  c  c  c  c }
		\hline
			 &    & $\langle b \rangle/\langle s^2 \rangle$  &   &  \\ \hline
%			\hline
			Topology & HP   & EPP(4\%)  & EPP(9\%)  & EPP(13\%) \\ 
			\hline
			Linear	&	0.409	&	0.407	&	0.395	&	0.401		\\ 
			Ring	& 	0.335	&	0.329	&	0.327	&	0.331		\\ 
			Trefoil	&	0.316	&	0.314	&	0.319	&	0.316		\\ \hline
%			\hline
			 &    & $\langle c \rangle/\langle s^2 \rangle$  &   &  \\ \hline
			Linear	&	0.066	&	0.066	&	0.073	&	0.073		\\ 
			Ring	& 	0.096	&	0.097	&	0.106	&	0.104		\\ 
			Trefoil	&	0.102	&	0.105	&	0.093	&	0.094		\\ \hline
%			\hline
			 &    &  $\langle \kappa^2 \rangle$  &   &  \\ \hline
			Linear	&	0.412	&	0.401	&	0.379	&	0.380		\\ 
			Ring	& 	0.263	&	0.230	&	0.229	&	0.234		\\ 
			Trefoil	&	0.238	&	0.203	&	0.211	&	0.206		\\ \hline
%			\hline
			 &    & $\langle p \rangle$   &   &  \\ \hline
			Linear	&	0.101	&	0.096	&	0.093	&	0.085		\\ 
			Ring	& 	0.028	&	0.021	&	0.024	&	0.026		\\ 
			Trefoil	&	0.028	&	0.025	&	0.033	&	0.029		\\ \hline
		\end{tabular}
		\label{table: b1}
	\end{center}
\end{table}

\begin{figure}[ht!]
	\begin{center}
		\includegraphics[width=0.47\textwidth]{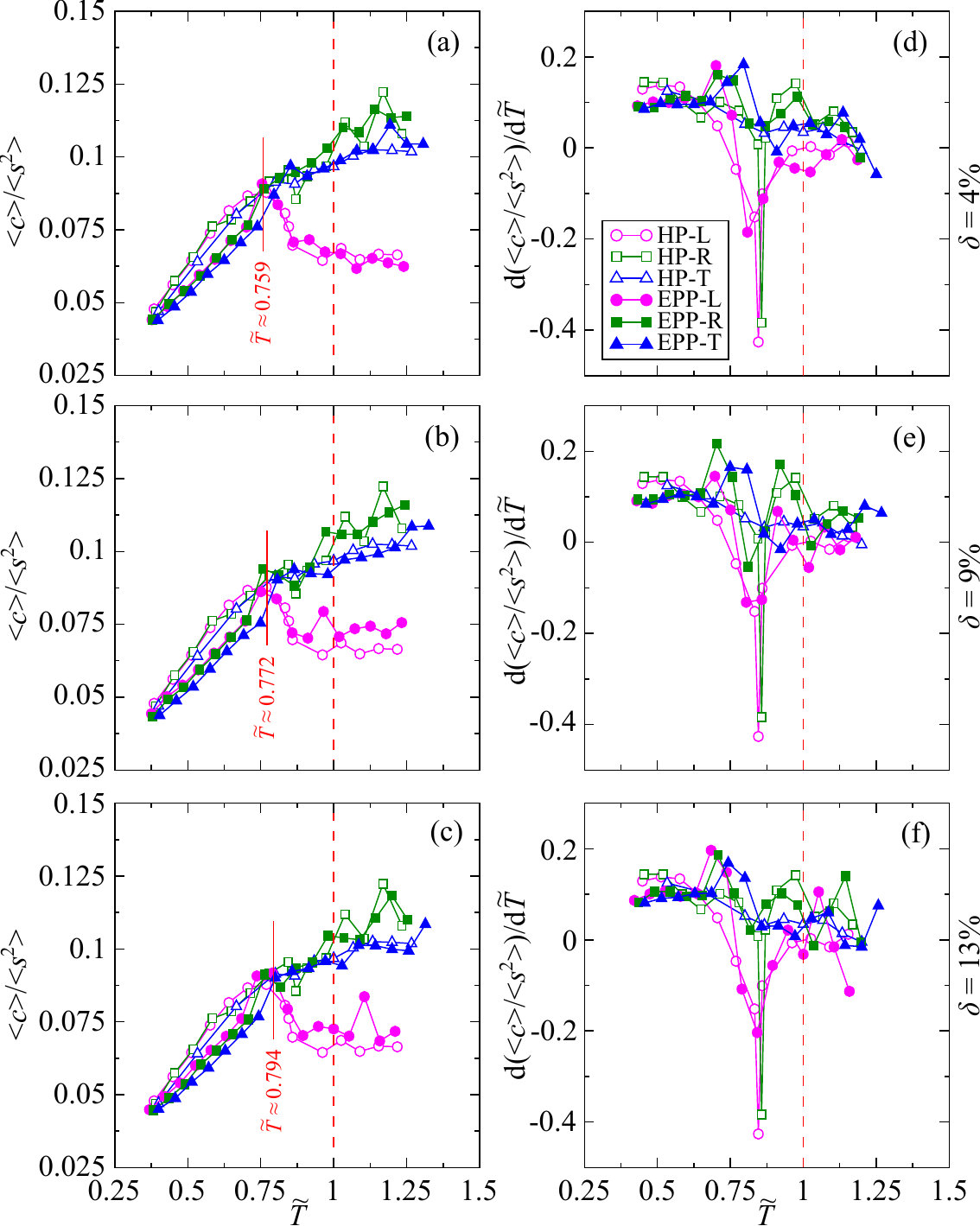}
		\caption{The plots of acylindricity, $\langle c \rangle/\langle s^2 \rangle$ as a function of $\tilde{T}$ for different topology are displayed in (a)-(c), and their temperature derivaties are in (d)-(f) respectively.}
		\label{fig: c}	
	\end{center}
\end{figure}

%\begin{table}
%	\begin{center}
%		\caption{$\langle c \rangle/\langle s^2 \rangle$ at $\tilde{T} = 1$}
%		\begin{tabular}{ c  c  c  c  c }
%			\hline
%			Topology & HP   & EPP(4\%)  & EPP(9\%)  & EPP(13\%) \\ \hline
%			Linear	&	0.066	&	0.066	&	0.073	&	0.073		\\ 
%			Ring	& 	0.096	&	0.097	&	0.106	&	0.104		\\ 
%			Trefoil	&	0.102	&	0.105	&	0.093	&	0.094		\\ \hline
%		\end{tabular}
%		\label{table: c1}
%	\end{center}
%\end{table}

\begin{figure}[ht!]
	\begin{center}
		\includegraphics[width=0.47\textwidth]{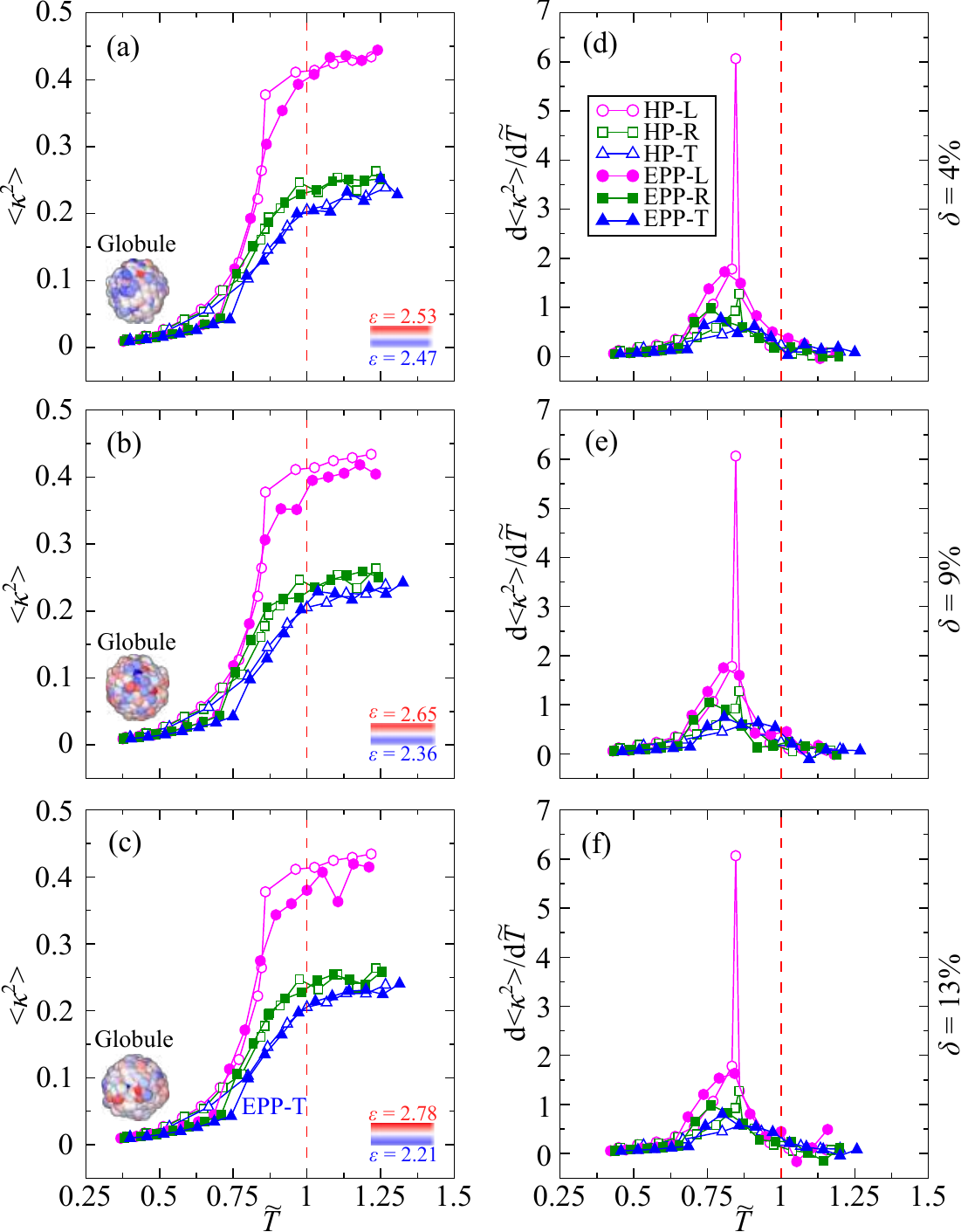}
		\caption{The plots of relative shape anisotropy, $\langle \kappa^2 \rangle$ as a function of $\tilde{T}$ for different topology are displayed in figure (a)-(c) at $\delta$ = 4, 9 and 13\% respectively. The derivates for the same are in (d)-(f) respectively.}
		\label{fig: k}	
	\end{center}
\end{figure}

Yet another quantity that we use in our investigation is acylindricity:
\begin{equation}
\langle c \rangle = \langle \lambda_2^2 \rangle - \langle \lambda_1^2 \rangle,
\end{equation} which measures how the conformation of the polymer deviates from the cylindrical shape. For a perfect cylinder, its value is zero. Figure~\ref{fig: c} displays the dependence of $\langle c \rangle/\langle s^2 \rangle$ on $\tilde{T}$ for different topology at various $\delta$, as well as their respective temperature derivatives. Linear chains assume cylindrical shape more than rings and trefoils at high $\tilde{T}$. This nature becomes quite distinct at $\tilde{T}$ = 0.759, 0.772 and 0.794 for EPPs $\delta$ = 4, 9 and 13\% respectively. These are reflected in their temperature derivative curves. Above these temperatures, rings and trefoils exhibit similar behaviour as $\langle c \rangle/ \langle s^2 \rangle$ increases and saturates at approximately 0.1. Trefoil conformations are more cylinderical than that of the rings. For the linear chains, saturation happens at a lower value of $\langle c \rangle/ \langle s^2 \rangle \approx$ 0.06. Furthermore, EEPs are less cylindrical than their homopolymer counterparts in high temperature regime, and that this trend increases with increasing polydispersity. The values of $\langle c \rangle/\langle s^2 \rangle$ at $\tilde{T} = 1$ are summarised in table~\ref{table: b1}. Below $\tilde{T}$ = 0.759, 0.772 and 0.794 in figure~\ref{fig: c}(a), (b) and (c) respectively, the effect of energy polydispersity is observed till $\tilde{T}$ = 0.375. In these regions, homopolymers are grouped together and so are the EPPs. Finally, they merge at $\tilde{T} \approx 0.40$. The general behavior is that, as temperature decreases, the rings and trefoil knots show a monotonic decrease in value of $\langle c \rangle/\langle s^2 \rangle$ (becomes more cylindrical), while the linear chains shows a non-monotonic behavior.  

\begin{figure}[ht!]
	\begin{center}
		\includegraphics[width=0.47\textwidth]{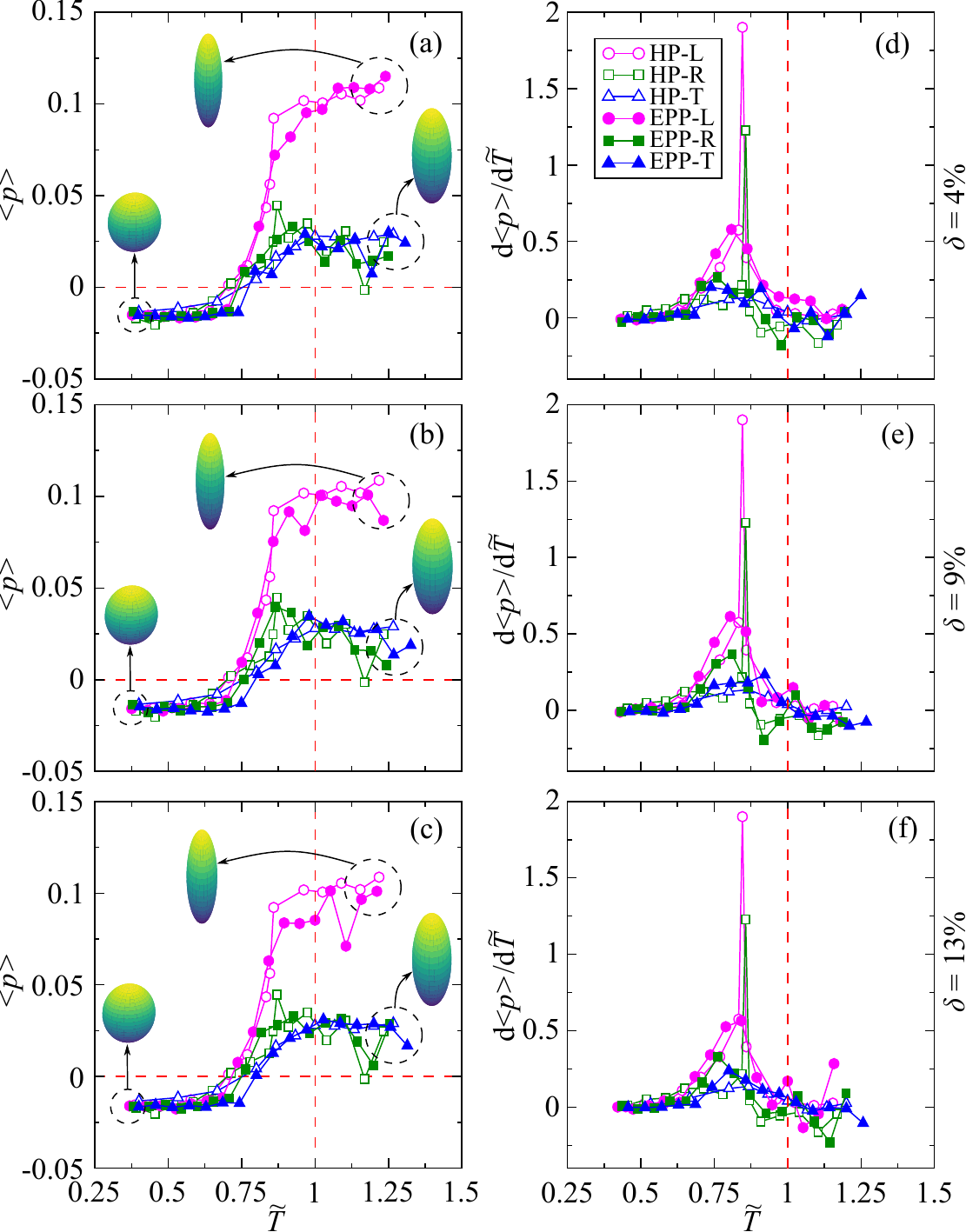}
		\caption{The plots of prolateness parameter $\langle p \rangle$ as a function of temperature for different topology at $\delta$ = 4, 9 and 12\% are shown in figure (a)-(c) and the temperature derivatives are in (d)-(f) respectively.}
		\label{fig: p}	
	\end{center}
\end{figure}

\begin{figure*}[ht!]
	\begin{center}
		\includegraphics[width=1\textwidth]{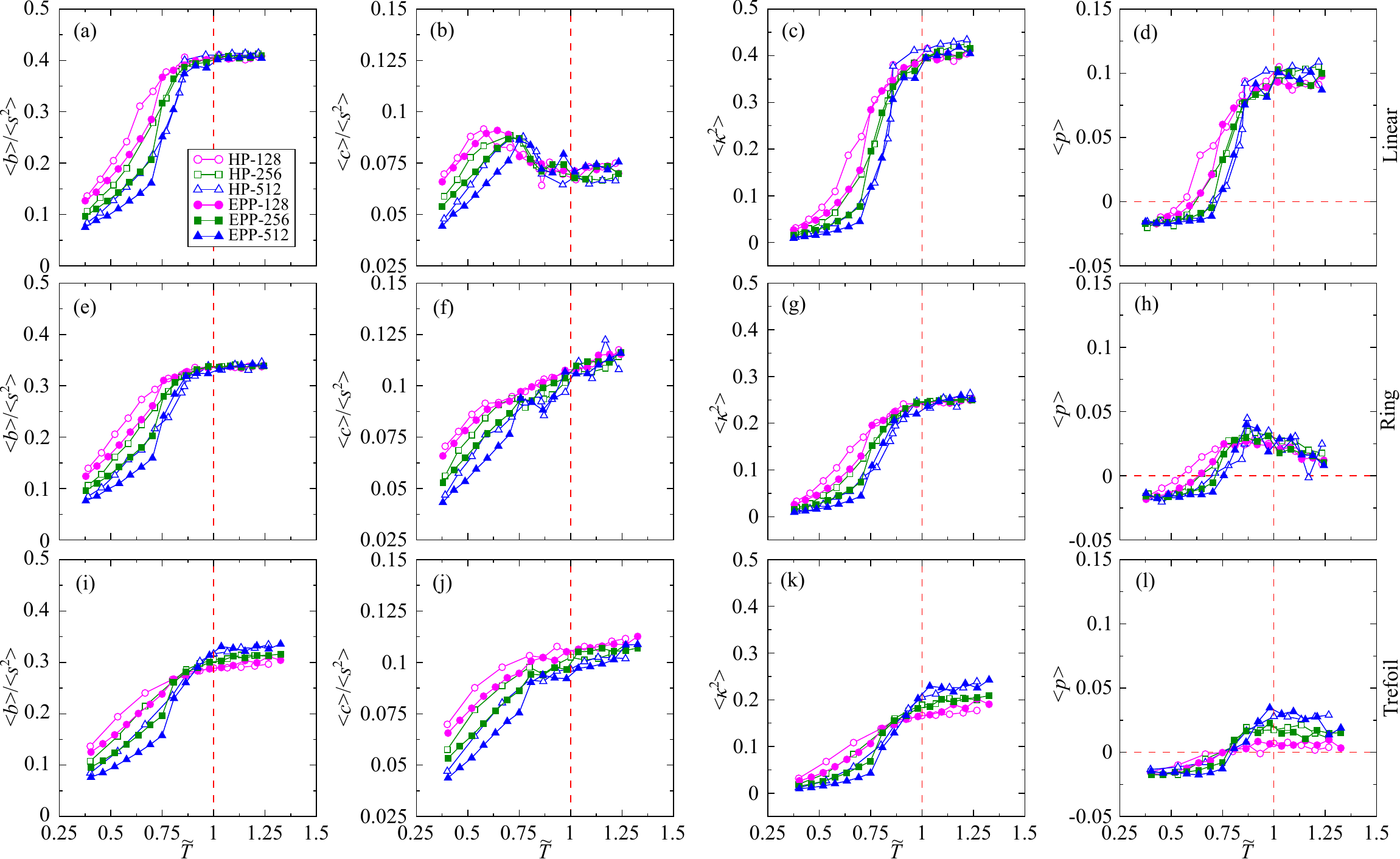}
		\caption{The asphericity, $\langle b \rangle/\langle s^2 \rangle$, cylindricity, $\langle c \rangle/\langle s^2 \rangle$, relative shape anisotropy, $\langle \kappa^2 \rangle$ and prolateness parameter, $\langle p \rangle$, as a function of temperature, $\tilde{T}$, for linear (a)-(d), simple ring (e)-(h), and trefoil knot (i)-(l) at $\delta$ = 9\%.}
		\label{fig: Nd9}	
	\end{center}
\end{figure*}

The relative shape anisotropy is defined as
\begin{equation}
\begin{aligned}
\langle \kappa^2 \rangle &= 1 - 3\left\langle \frac{\lambda_1^2 \lambda_2^2 + \lambda_2^2 \lambda_3^2 + \lambda_3^2 \lambda_1^2}{s^4} \right\rangle \\
&= \left\langle\frac{b^2 + 3c^2/4}{s^4} \right\rangle,
\end{aligned}
\end{equation} in which $ \langle s^2 \rangle = \langle R_g^2 \rangle = \langle \lambda_1^2 + \lambda_2^2 + \lambda_3^2  \rangle $. Its value is 1 for rod-like molecules and becomes 0 with those with perfect symmetry. The variation of $\langle \kappa^2 \rangle$ as a function of $\tilde{T}$, and their temperature derivatives are displayed in figure~\ref{fig: k} for different topology and polydispersity. At high temperature, the trefoils, irrespective of polydispersity, have the smallest values, and so, they possess the most symmetrical conformations, which are followed by rings. The deviation from the symmetric nature in linear chains becomes large at $\tilde{T} \approx$ 0.65. It is reflected in the temperature derivative curves (see figure~\ref{fig: k}(d)-(f)). This deviation continues till saturation is achieved at $\langle \kappa^2 \rangle \approx$ 0.4. The saturation for the rings and the trefoils are observed at $\langle \kappa^2 \rangle \approx$ 0.25 and 0.24 respectively. Moreover, higher polydispersity results in more symmetric nature for the linear chains. The difference in the other two topologies are not highly significant. Their values at $\tilde{T}$ = 1 are summarised in table~\ref{table: b1}. In the reduced temperature range of 0.5 to 0.7, the grouping of the homopolymer and EPP curves are seen. And at very low temperature, they all merge together as all of them form compact globules.

%\begin{table}
%	\begin{center}
%		\caption{$\langle \kappa^2 \rangle$ at $\tilde{T} = 1$}
%		\begin{tabular}{ c  c  c  c  c }
%			\hline
%			Topology & HP   & EPP(4\%)  & EPP(9\%)  & EPP(13\%) \\ \hline
%			Linear	&	0.412	&	0.401	&	0.379	&	0.380		\\ 
%			Ring	& 	0.263	&	0.230	&	0.229	&	0.234		\\ 
%			Trefoil	&	0.238	&	0.203	&	0.211	&	0.206		\\ \hline
%		\end{tabular}
%		\label{table: k1}
%	\end{center}
%\end{table}

%\begin{table}
%	\begin{center}
%		\caption{$\langle p \rangle$ at $\tilde{T} = 1$}
%		\begin{tabular}{ c  c  c  c  c }
%			\hline
%			Topology & HP   & EPP(4\%)  & EPP(9\%)  & EPP(13\%) \\ \hline
%			Linear	&	0.101	&	0.096	&	0.093	&	0.085		\\ 
%			Ring	& 	0.028	&	0.021	&	0.024	&	0.026		\\ 
%			Trefoil	&	0.028	&	0.025	&	0.033	&	0.029		\\ \hline
%		\end{tabular}
%		\label{table: p1}
%	\end{center}
%\end{table}

Finally, we study the prolateness parameter defined as
\begin{equation}
\langle p \rangle = \left\langle \frac{\left( 2\lambda_1 - \lambda_2 - \lambda_3 \right)\left( 2\lambda_2 - \lambda_1 - \lambda_3 \right)\left( 2\lambda_3 - \lambda_1 - \lambda_2 \right)}{2 \left( \lambda_1^2 + \lambda_2^2 + \lambda_3^2 -\lambda_1 \lambda_2 - \lambda_2 \lambda_3 - \lambda_1 \lambda_3 \right)^{3/2}} \right\rangle.
\end{equation} For a perfectly prolate shape, $\langle p \rangle = 1$, and for a perfectly oblate shape, $\langle p \rangle = -1$. In figure~\ref{fig: p}, the prolateness parameter as a function of $\tilde{T}$, and its derivatives are displayed for different topologies. At high temperature, linear chains are quite prolate compared to rings and trefoils. Linear chains saturate at $\langle p \rangle \approx$ 0.1, while the others do so at $\langle p \rangle \approx$ 0.02. The values of $\langle p \rangle$ at $\tilde{T}$ = 1 are summarised in table~\ref{table: b1}. Between $\tilde{T}$ = 0.75 and 1, we observe that the ring conformations are more prolate than that of the trefoils. Below $\tilde{T} \approx 0.75$, all the curves merge, and they become oblate, which is indicated by the horizontal dashed line. The effect of polydispersity is prominently seen in the linear chains at temperatures below $\tilde{T} = 0.75.$ The EPP linear chins are less prolate at this tempearture range. No significant differences is observed for the rings and the trefoils. In general, as the temperature decreases the prolate to oblate transition is more rapid for trefoils, followed by the rings.  

%%%%%%%%%%%%%%%%%%%%%%%%%%%%%%%%%%%%%%%%%%%%%%%%%%
%%%%%%%%%%%%%%%%%%%%%%%%%%%%%%%%%%%%%%%%%%%%%%%%%%
\subsection{Effect of chain length} 
\label{sec: chain-length}

The variations of asphericity, $\langle b \rangle/\langle s^2 \rangle$, cylindricity, $\langle c \rangle/\langle s^2 \rangle$, relative shape anisotropy, $\langle \kappa^2 \rangle$ and prolateness parameter, $\langle p \rangle$, as a function of reduced temperature, $\tilde{T}$,  are displayed in figure~\ref{fig: Nd9} for different chain length ($N$ = 128, 256 and 512) at $\delta$ = 9\% (See SI for $\delta$ = 4 and 13\%). A general behaviour thus observed is that the effect of chain length is significant for $T < \tilde{T}$ = 1. Similar observation is also seen in a different study where log-normal distribution is considered at low polydispersity index (1\%)\cite{Vilip2021}. Only linear chains with $N$ = 64 and 512 are considered in that study.

\begin{figure*}[ht!]
	\begin{center}
		\includegraphics[width=1\textwidth]{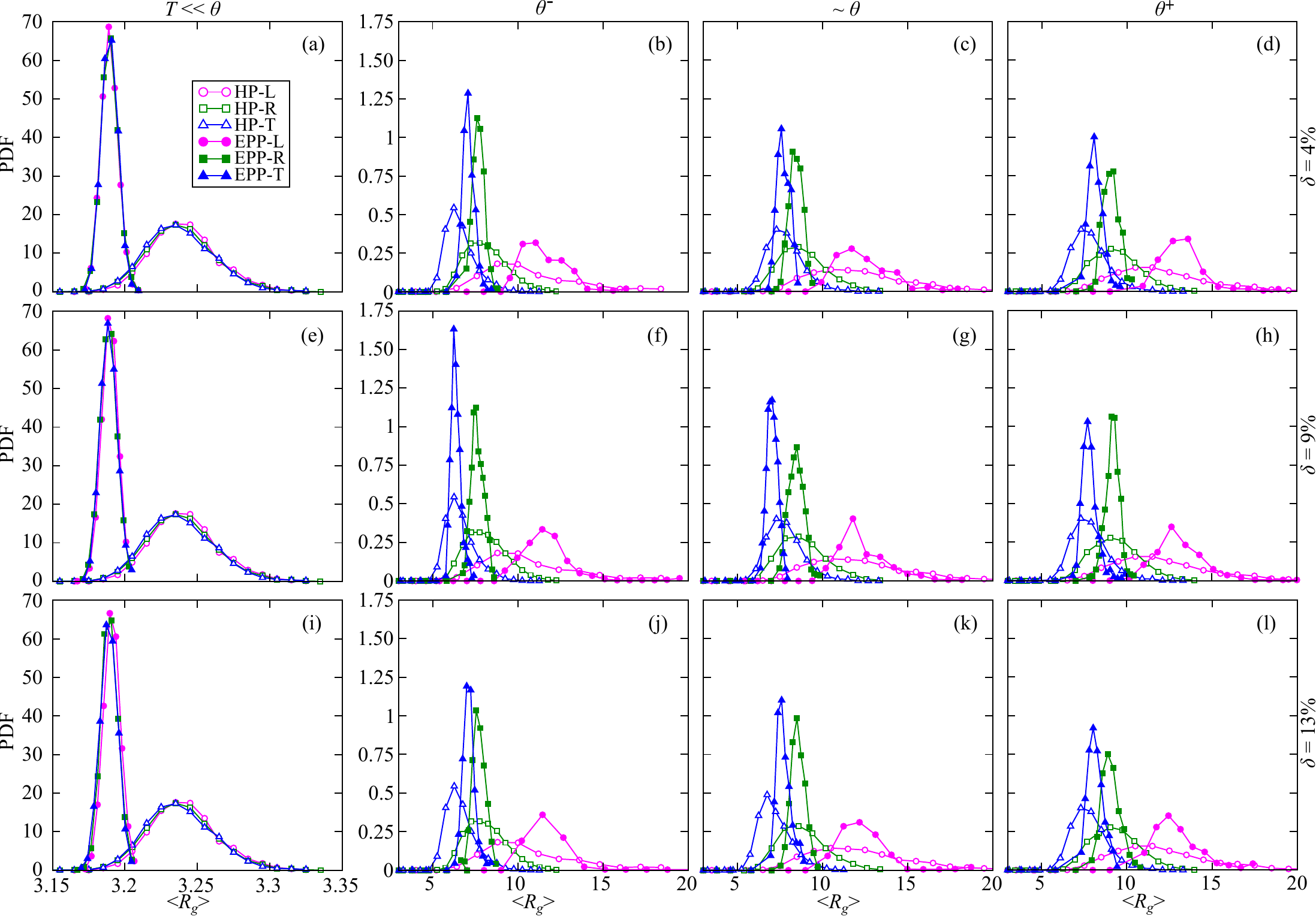}
		\caption{The PDF of $\langle R_g \rangle$ at $T << \theta$, $\theta^-$(below 5\%), approximately $\theta$, and  $\theta^+$(above 5\%) are plotted at $\delta$ = 4\% (a)-(d), 9\% (e)-(h), and 13\% (i)-(l). All polymers have a chain length of $N$ = 512 monomers. Open symbols indicate homopolymers and closed ones are for EPPs.}
		\label{fig: Pdf-rg}	
	\end{center}
\end{figure*}

The effect of chain length on how $\langle b \rangle/\langle s^2 \rangle$ depends on $\tilde{T}$ is shown in figure~\ref{fig: Nd9}(a), (e), (i) for linear, ring and trefoil respectively. Linear and ring polymers are less symmertical at high temperature and their curves  merge at $\tilde{T} \geq 0.88$. As for the trefoil  knotes, longer the chain, the more it deviates from spherical symmetry. Below $\tilde{T} \approx 0.88$, the effect of chain length is quite dominant. Shorter chains are less spherical at the same $\tilde{T}$, and that homopolymers of the same chain length are even more so.

In figure~\ref{fig: Nd9}(b), (f) and (j), the variation of $\langle c \rangle/\langle s^2 \rangle$ as $\tilde{T}$ changes is displayed for different chain lenght and different topology. At low temperature, the trefoils possess the most cylindrical nature, followed by the rings. They deviate from this nature as temperature increases. In figure~\ref{fig: Nd9}(f) and (j), the curves tend to merge at temperatures beyond $\tilde{T}$ = 1. However, in figure~\ref{fig: Nd9}(b), the linear chains become  more cylindrical at temperatures above 0.62. The trends thus observed for all topologies are the same for the homopolymers and the EPPs, but the EPPs have smaller values of $\langle c \rangle/\langle s^2 \rangle$ at the same temperature when compared to their respective topologies.

We also study the effect of chain length on how $\langle \kappa^2 \rangle$ changes with $\tilde{T}$ in figure~\ref{fig: Nd9}(c), (g) and (k). The dependence of chain length is quite visible in the range of $\tilde{T} \approx$ 0.5 to 0.88. The shorter chains have larger values of $\langle \kappa^2 \rangle$ meaning they posses less symmetry. Above $\tilde{T} \approx$ 0.88, polymers of all topologies have similar symmetry, except for the trefoil rings. The EPPs have higher values of $\langle \kappa^2 \rangle$ compared to their homopolymer counterparts. Unlike that of $\langle b \rangle/\langle s^2 \rangle$, and $\langle c \rangle/\langle s^2 \rangle$, polymers of all chain lenghts have values of $\langle \kappa^2 \rangle$ very close to one another at a very low temperature of $\tilde{T} \approx$ 0.4.

The prolatelenss parameter, $\langle p \rangle$, for different chain length are also investigated for linear, ring and trefoil polymers as shown in figure~\ref{fig: Nd9}(d), (h) and (l) respectively. This trend is similar to that of $\langle \kappa^2 \rangle$. At low temperatures, all topological polymers are oblate. The oblate to prolate transition is indicated by the horizontal dashed line. Shorter chains show this transition at lower temperature compared to the longer chains. Moreover, the homopolymers are also more prolate compared to their EPP counterparts. In figure~\ref{fig: Nd9}(d), the curves merge at $\langle p \rangle \approx$ 0.1 above $\tilde{T} \approx$ 0.88. As for the simple rings in figure~\ref{fig: Nd9}(h), they are most prolate at $\tilde{T} \approx$ 0.88, above which the prolate nature decreases. Figure~\ref{fig: Nd9}(l) is for the trefoil rings, and shows that the longest chains are the most prolate, and vice versa for shorter chains at high temperatures. However, the dependence of chain length vanishes at low temperature. The plots for the same at $\delta$ = 4\% and 13\% are in SI.

%%%%%%%%%%%%%%%%%%%%%%%%%%%%%%%%%%%%%%%%%%%%%%%%%%
%%%%%%%%%%%%%%%%%%%%%%%%%%%%%%%%%%%%%%%%%%%%%%%%%%

\subsection{Chain size distribution and effect of topology} 
\label{sec: probability-distribution}

The size of a polymer chain is measured by the ensemble averaged radius of gyration, $\langle R_g \rangle$. The chain size distribution is calculated for a single chain length, $N$ = 512, for different topology at $\delta$ = 4, 9 and 13\%. They are done so at four different temperatures: (i) $T << \theta$, (ii) $T \approx \theta$ (iii) $T = \theta^- = \theta\left(1 - a\right)$, and (iv) $T = \theta^+ = \theta\left(1 + a\right)$ in which $a$ = 0.05. The probability density function (PDF) of $\langle R_g \rangle$ for different topologies are displayed in figure~\ref{fig: Pdf-rg}.

In figure~\ref{fig: Pdf-rg}(a), (e), and (i), the chain size distributions are studied at $\delta$ = 4, 9, and 13\% respectively, at a low temperature ($T \approx 1.5 << \theta$), in which the chains of all topologies assume a globular state. No significant effect of topology is seen at this temperature. However, the difference between the EPPs and the homopolymers are quite evident. The EPPs form globules far more compact than their corresponding homopolymers, and as such the distributions of EPPs for all values of $\delta$ are narrower and have smaller values of $\langle R_g \rangle$. The influence of polydispersity is also negligible at this temperature. 

In all the remaining plots, i.e.,figure~\ref{fig: Pdf-rg}(b)-(d), (f)-(h), (j)-(l), the unknotted and the knotted rings have distributions narrower than that of the linear chains. The linear chains have larger $\langle R_g \rangle$. And the ones with the narrowest distributions with the smallest values of $\langle R_g \rangle$ are the trefoils due to its topological constraints. In general, the peaks for EPPs are narrower than that of their homopolymers. Moreover, as $\langle R_g \rangle$ gets larger with increasing temperature, the distributions become broader as well. 

%\clearpage

%%%%%%%%%%%%%%%%%%%%%%%%%%%%%%%%%%%%%%%%%%%%%%%%%% 
%%%%%%%%%%%%%%%%%%%%%%%%%%%%%%%%%%%%%%%%%%%%%%%%%%

\section{Conclusion}
\label{sec: summary}

By considering a coarse-grained bead-spring polymer model in an implicit solvent, we define EPPs whose monomeric interaction energies are derived from a Gaussian distribution. The EPPs are characterised by the polydispersity index, $\delta$ = SD/mean (4, 9, and 13\%). First, we determine the $\theta$-temperature, defined as the theperature where $\textit{g-factor}$ is unity, for linear chains, unknotted and knotted (trefoil) rings as shown in figure~\ref{fig: xi-vs-T}. The effect of topology on $\theta$ is that it is highest for the linear chains and the lowest for the trefoil knot. Introduction of topological contraints results in achieving $\theta$ at a lower temperature for both homopolymers and EPPs.

Next, the topological dependence on the instantaneous shapes of EPPs are discussed in details. Asphericity, $\langle b \rangle/\langle s^2 \rangle$, acylindricity, $\langle c \rangle/\langle s^2 \rangle$, relative shape anisotropy, $\langle \kappa^2 \rangle$, and prolateness parameter, $\langle p \rangle$, are studied as a function of reduced temperature, $\tilde{T} = T/\theta$. Trefoils possess higher spherical symmetry, followed by rings. This is true for acylindricity as well at low temperature upto around $\tilde{T} \approx$ 0.759, 0.772 and 0.794 at $\delta$ = 4, 9 and 13\% respectively, above which the linear chains have lower values of  $\langle c \rangle/\langle s^2 \rangle$ indicating more cylindircal symmetry. All polymers are oblate at low temperatures, and with increasing temperature, they transform into prolate structures. The linear chains become the most prolate, while both types of rings are less prolate with similar positive values of $\langle p \rangle$. Furthermore, we investigate the chain length dependence on the instantaneous shapes of linear, ring and trefoil polymers by considering $N$ = 128, 256 and 512. For linear chains and unknotted rings, the effect of chain length becomes dominant below $\tilde{T} \approx 1$. The longer chains are more symmetric, whether it be a spherical  or a cylindrical symmetry. Above it, the curves representing a particular shape parameter merge together. On the other hand, the trefoil rings show a peculiar behaviour. As $\tilde{T} > 0.88$, the longest trefoil rings have the greatest deviation from spherical symmetry and highest value of $\langle \kappa^2 \rangle$. Longer chains possess oblate shape upto $\tilde{T} \approx$ 0.75, and shorter chains are oblate upto a less $\tilde{T}$.  Similarly, the EPPs of all chain lenghts assume more compact structures as compared to that of the homopolymer counterparts. 
Finally, the effect of topology on the chain size distribution is investigated. At very low temperature, the size distribution is the same for all the EPPs and have narrow distribution, also the influence of polydispersity is negligible.

%\clearpage
%%%%%%%%%%%%%%%%%%%%%%%%%%%%%%%%%%%%%%%%%%%%%%%%%%%%%%%%
%%%%%%%%%%%%%%%%%%%%%%%%%%%%%%%%%%%%%%%%%%%%%%%%%%%%%%%%

\begin{acknowledgements}
T.~Vilip acknowledges fruitful discussions with M.~Premjit, J.~Pame, and U.~Somas. LSS's research was supported in part by DST-INSPIRE Faculty Award (grant number: DST/INSPIRE/04/2015/001914).
\end{acknowledgements}

%%%%%%%%%%%%%%%%%%%%%%%%%%%%%%
%%%%%%%%%%%%%%%%%%%%%%%%%%%%%%

%

\begin{thebibliography}{}

	\bibitem{Karger-Kocsis} J. Karger-Kocsis, \textit{Polypropylene: an AZ reference}, Springer Science \& Business Media {\bf 2012}. 
	
	\bibitem{Ren} J.M. Ren, T.G. McKenzie, Q. Fu, E.H. Wong, J. Xu, Z. An, S. Shanmugam, T.P. Davis, C. Boyer, G.G. Qiao \textit{Chem. Rev.}, {\bf 2016}, {\it 116}, 6743.
	
	\bibitem{DeKeer} L. De Keer, K.I. Kilic, P.H. Van Steenberge, L. Daelemans, D. Kodura, H. Frisch, K. De Clerck, M.-F. Reyniers, C. Barne-Kowollik, R.H. Dauskardt, D.R. D’hooge \textit{Nat. Mater.}, {\bf 2021}, {\it 20}, 1422.
	
	\bibitem{mcleish2002} T.~McLeish, {\it Science}, {\bf 2002}, {\it 297}, 2005. 
	
	\bibitem{polymerpoulous207} G.~Polymeropoulos, G.~Zapsas, K.~Ntetsikas, P.~Bilalis, Y.~Gnanou, N.~Hadjichristidis {\it Macromolecules}, {\bf 2017}, {\it 50}, 1253. 
	
	\bibitem{halverson2014} J.D. Halverson, J. Smrek, K. Kremer, A. Y. Grosberg {\it Rep.~Prog.~Phys.}, {\bf 2014}, {\it 77}, 022601. 
	
	\bibitem{shagolsem2018} L.S.~Shagolsem, {\it J.~Phys.~Chem.~B}, {\bf 2018}, {\it 122}, 1306. 
	
	\bibitem{Freifelder} D. Freifelder, A.K. Kleinschmidt, R.L. Sinsheimer \textit{Science}, {\bf 1964}, {\it 146}, 254.
	
	\bibitem{Wasserman1} S.A. Wasserman, N.R. Cozzarelli \textit{Science}, {\bf 1986}, {\it 232}, 951.
	
	\bibitem{Champoux} J.J. Champoux \textit{Annu. Rev. Biochem.}, {\bf 2001}, {\it 70}, 369.
	
	\bibitem{Meluzzi} D. Meluzzi, D.E. Smith, G. Arya \textit{Annu. Rev. Biophys.}, {\bf 2010}, {\it 39}, 349.
	
	\bibitem{Yeates} T.O. Yeates, T.S. Norcross, N.P. King \textit{Curr. Opin. Chem. Biol.}, {\bf 2007}, {\it 11}, 595.
	
	
	\bibitem{Rybekov} V.V. Rybenkov, N.R. Cozzarelli, A.V. Vologodskii \textit{Proc. Natl. Acad. Sci. U. S. A.}, {\bf 1993} {\it 90}, 5307.
	
	\bibitem{Shaw} S.Y. Shaw, J.C. Wang \textit{Science}, {\bf 1993}, {\it 260}, 533.
	
	\bibitem{Buck} D. Buck, E. Flapan \textit{J. Mol. Biol.}, {\bf 2007} {\it 374}, 1186.
	
	\bibitem{Lui} Z. Liu, R.W. Deibler, H.S. Chan, L. Zechiedrich \textit{Nucleic Acids Res.}, {\bf 2009}, {\it 37}, 661.
	
	\bibitem{Stark} W.M. Stark, M.R. Boocock \textit{J. Mol. Biol.}, {\bf 1994}, {\it 239}, 25.
	
	\bibitem{Wasserman} S.A. Wasserman, J.M. Dungan, N.R. Cozzarelli \textit{Science}, {\bf 1985}, {\it 229}, 171.
	
	%\bibitem{Janse} Janse van Rensburg, E. J. (2009). \textit{Thoughts on lattice knot statistics.} Journal of mathematical chemistry, 45(1), 7-38.
	
	%\bibitem{Grosberg} Grosberg, A. Y. (2009). \textit{A few notes about polymer knots.} Polymer Science Series A, 51(1), 70-79.
	
	%\bibitem{Bao} Bao, X. R., Lee, H. J., \& Quake, S. R. (2003). \textit{Behavior of complex knots in single DNA molecules.} Physical review letters, 91(26), 265506.
	
	
	\bibitem{Ernst} C. Ernst, D. Sumners \textit{Math. Proc. Camb. Philos. Soc.}, {\bf 1990}, {\it 108}, 489
	
	\bibitem{Shimokawa} K. Shimokawa, K. Ishihara, I. Grainge, D.J. Sherratt, M. Vazquez \textit{Proc. Natl. Acad. Sci. U. S. A.}, {\bf 2013}, {\it 110}, 20906.
	
	\bibitem{Seeman} N.C. Seeman \textit{Annu. Rev. Biochem.}, {\bf 2010} {\it 79}, 65.
	
	\bibitem{kapnistos2008} M.~Kapnistos, M.~Lang, D.~Vlassopoulos, W.~Pyckhout-Hintzen, D.~Richter, D.~Cho, T.~Chang, M.~Rubinstein, {\it Nat. Mater.}, {\bf 2008}, {\it 7}, 997.

%%	\bibitem{halverson2012} J.D.~Halverson, G.S.~Grest, A.Y.~Grosberg, K.~Kremer, {\it Phys. Rev. Lett.} {\bf 2012}, 108, 038301. 

	\bibitem{richter2015} D.~Richter, %S.~Goo{\ss}en, A.~Wischnewski, {\it Soft Matter}, {\bf 2015}, {\it 11}, 8535−8549.
	
	\bibitem{Kuhn} W.~Kuhn \textit{Kolloid-Zeitschrift}, {\bf 1934}, {\it 68}, 2.
	
	\bibitem{Solc} K.~\v{S}olc \textit{J. Chem. Phys.}, {\bf 1971}, {\it 55}, 335. 
	
	\bibitem{Haber} C. Haber, S.A. Ruiz, D. Wirtz \textit{Proc. Natl. Acad. Sci. U. S. A.}, {\bf 2000}, {\it 97}, 10792.
	
	\bibitem{Abernathy} F.H. Abernathy, J.R. Bertschy, R.W. Chin, D.E. Keyes  \textit{J. Rheol.}, {\bf 1980}, {\it 24}, 647.
	
	\bibitem{Rohs} R. Rohs, S.M. West, A. Sosinsky, P. Liu, R.S. Mann, B. Honig \textit{Nature}, {\bf 2009}, {\it 461}, 1248.
	
	\bibitem{Steinhauser} M.O. Steinhauser \textit{J. Chem. Phys.}, {\bf 2005}, {\it 122}, 094901.
	
	\bibitem{Quake} S.R. Quake \textit{Phys. Rev. Lett.}, {\bf 1994}, {\it 73}, 3317.
	
	\bibitem{Grosbergg} A.Y. Grosberg, A. Feigel, Y. Rabin \textit{Phys. Rev. E}, {\bf 1996}, {\it 54}, 6618.
	
	\bibitem{Derrida1} B. Derrida \textit{Phys. Rev. Lett.}, {\bf 1980}, {\it 45}, 79.
	
	\bibitem{Derrida2} B. Derrida \textit{Phys. Rev. B}, {\bf 1981}, {\it 24}, 2613.
	
	\bibitem{Pande} V.S. Pande, A.Y. Grosberg,C. Joerg, T. Tanaka \textit{Phys. Rev. Lett.}, {\bf 1996}, {\it 76}, 3987.
	
	\bibitem{Shakhnovich} E.I. Shakhnovich, A.M. Gutin \textit{Biophys. Chem.}, {\bf 1989}, {\it 34}, 187.
	
	\bibitem{Vilip2021} T.V. Singh, L.S. Shagolsem \textit{Macromol. Symp.}, {\bf 2021}, {\it 399}, 2100002.
	
	\bibitem{vilip2022} T.V. Singh, L.S. Shagolsem {\it (Preprint) arXiv:1909.04478}, submitted: July 2022.
	
	\bibitem{Grest} G.S. Grest, K. Kremer (1986). \textit{Phys. Rev. A}, {\bf 1986}, {\it 33}, 3628.
	
	\bibitem{Berthelot} D. Berthelot {\it Comptes. Rendus. Acad. Sci. Paris}, {\bf 1889}, {\it 126}, 1703
	
	\bibitem{Kremer} C.F. Abrams, K. Kremer \textit{J. Chem. Phys.}, {\bf 2001}, {\it 115}, 2776.
	
	\bibitem{Allen} M.P. Allen, D.J. Tildesley, \textit{Computer Simulation of Liquids}, Oxford University Press, Oxford, {\bf 2017}.
	
	\bibitem{Frenkel} D. Frenkel, B.Smit \textit{Understanding Molecular Simulation: From Algorithms to Applications}, Academic Press, Cambridge, {\bf 2001}.
	
	\bibitem{Plimpton} S. Plimpton (1995) \textit{J. Comp. Phys.}, {\bf 1995}, {\it 117}, 1.
	
	\bibitem{Grosberg_book} A.Y. Grosberg, A.R. Khokhlov \textit{Statistical Physics of Macromolecules.} AIP, Woodbury, NY, {\bf 1994}
	
	\bibitem{Cates} M.E. Cates, Deutsch, J. M. {\it Journal de physique.}, {\bf 1986}, {\it47}, 2121.
	
	\bibitem{Zimm} B.H. Zimm, W.H. Stockmayer \textit{J. Chem. Phys.}, {\bf 1949}, {\it 17}, 1301.
	
	\bibitem{Suzuki} J. Suzuki, A. Takano, Y. Matsushita. \textit{J. Chem. Phys.}, {\bf 2013}, {\it 138}, 024902.
	
	\bibitem{Narros} A. Narros, A.J. Moreno, C.N. Likos. {\it Macromolecules}, {\bf 2013}, {\it 46}, 3654.
	
	\bibitem{solc_stockmayer_1971} K. \v{S}olc, W.H. Stockmayer {\it J. Chem. Phys.}, {\bf 1971}, {\it 54}, 2981.
		
	\bibitem{solc1973} K. \v{S}olc, \textit{Macromolecules}, {\bf 1973}, {\it 6}, 378



%%%%%%%%%%


\end{thebibliography}
\end{document}